# Controlled Modulation of Depolarization in Laser Speckle

ABHIJIT ROY,[1] RAKESH K. SINGH,[2] MARUTHI M. BRUNDAVANAM[1,*]

[1]Department of Physics, Indian Institute of Technology Kharagpur, West Bengal 721302, India
[2]Applied and Adaptive Optics Laboratory, Department of Physics, Indian Institute of Space Science and Technology (IIST), Trivandrum, Kerala 695547, India
*Corresponding author: bmmanoj@phy.iitkgp.ernet.in

**A new technique based on superposition of two speckle patterns is proposed and demonstrated for controlled modulation of the spatial polarization distribution of the resultant speckle. It is demonstrated both theoretically and experimentally that controlled modulation of the spatial polarization distribution of laser speckle can be achieved by proper choice of the polarization states as well as the average spatial intensity of the constituent speckles. It is also shown that the proposed technique is useful to generate different speckle patterns with sinusoidal variation in their degree of polarization, which can be tuned from zero to unity. This technique can find applications in sensing, biomedical studies, and in determining the rotation of the electric field vector after passing through a scattering medium.**

Interferometer since its inception has been widely used for various purposes and different kinds of interferometers are developed for wide range of studies [1-4]. For instance, two beam or multi-beam interferometric technique has been used to study different optical and physical properties [5-8], vortex beam [9, 10], in holography [11] as well as in cosmology [1]. Recently, interferometry method is also applied to generate network of vortices in a partially coherent beam [12]. Another kind of interferometer known as polarization interferometer is employed to control the state of polarization of the resultant beam [13, 14]. The interferometric technique is also used to study the polarization singularities of a vector field [15] and to study the unfolding of an optical vortex while propagating through an anisotropic crystal [16]. The interference of a plane wave with a scattered random field is also helpful to investigate different properties of the random field as well as to study the surface roughness of the scattering medium [17, 18].

The propagation of coherent light through a random scattering medium results in a randomly distributed grainy intensity pattern, known as the speckle [19]. Speckle generated from a scattering medium, where single scattering event dominates, retains the polarization state of the input beam, whereas a multiple scattering medium spatially scrambles the polarization state of the input beam and make the speckle completely depolarized, known as the polarization speckle [20]. In case of a spatially uniformly polarized speckle, the polarization state is independent of the position whereas for a depolarized speckle, the polarization state changes randomly with the position on the speckle pattern, resulting in zero net polarization [17].

Apart from different interesting properties observed in speckle generated from a single scattering medium, the study of superposition of two different speckles also has gained attention because of its wide application in different studies. The speckle interferogram, obtained from the superposition of two speckles is useful in different kinds of object recovery through a random scattering medium, [21, 22] and in other different physical property [23] and non-invasive bio-medical studies [24]. It is observed that the superposition of two speckles, depending on their mutual correlation, changes the statistics of the resultant speckle [25]. In this case, the probability density function (PDF) of intensity of the resultant speckle has negative exponential distribution for fully correlated constituent speckle and with the decrease of mutual correlation, the PDF starts to deviate from the exponential nature [25], and this study has been extended to the superposition of N number of uncorrelated speckle patterns [26]. The intensity correlation function of the resultant speckle is also found to be changing with changes in one of the constituent speckle [27].

The polarization, another interesting property of the speckle, has been intensely studied for the last few years for fundamental and practical reasons. The polarization characteristics of a speckle is found to be carrying the nature of the scattering medium, whether it is a single scatterer or bulk scatterer [28]. The recently introduced complex degree of mutual polarization can be employed to differentiate the scattering media generating same polarization distributions [29]. In another study, the change of local polarization distribution or the micro-statistics of speckle with the input polarization is investigated for different degree of depolarization of the medium [30]. Recently, controlling the polarization of a scattered light is demonstrated by focusing a beam into a multiple scattering medium [31]. In another work, it is theoretically predicted that the interference of two unpolarized fields can produce completely polarized and depolarized field depending on the spatial position [32]. It is also shown that the correlation of the polarization elements can be manipulated using a spiral phase plate (SPP) and by different holographic techniques [33-35]. In another recent development, generation of partially coherent beams with different complex degrees of coherence is demonstrated experimentally [36]. Although different techniques are developed to control the coherence-polarization (CP) matrix elements and to change the polarization state of scattered light, the controlled modulation of the coherence-polarization property has been limited. The degree of polarization is fixed in all the reported works, except in a theoretical investigation [32]. In this letter, a new technique based on interferometric approach, without using any SPP or hologram, is proposed and experimentally demonstrated to control the coherence and spatial polarization distribution of the resultant speckle. It is also shown that the degree of polarization of the resultant speckle can be tuned from zero to unity in a sinusoidal manner. The theoretical background along with the experimental demonstration is presented.

Let us consider that the electric field vector of a spatially randomly polarized monochromatic object random field, $\mathbf{E_O}(\mathbf{r}, t)$ at the transverse observation plane $\mathbf{r}$ can be written in terms of its orthogonal polarization components $E_{Ox}(\mathbf{r}, t)$ and $E_{Oy}(\mathbf{r}, t)$ as

$$\mathbf{E_O}(\mathbf{r}, t) = E_{Ox}(\mathbf{r}, t)\,\hat{\mathbf{x}} + E_{Oy}(\mathbf{r}, t)\,\hat{\mathbf{y}} \quad (1)$$

where $\hat{\mathbf{x}}$, $\hat{\mathbf{y}}$ are the unit orthogonal polarization vectors and t is the time. The coherence-polarization property of the spatially randomly polarized field can be characterized employing either the CP matrix following Tervo *et al.* [37] or using the two-point intensity correlation function following the HBT approach [38] as

$$\gamma^2(\mathbf{r_1}, \mathbf{r_2}) = \frac{\mathrm{tr}[\Gamma^O(\mathbf{r_1},\mathbf{r_2})\,\Gamma^{O\dagger}(\mathbf{r_1},\mathbf{r_2})]}{\mathrm{tr}[\Gamma^O(\mathbf{r_1},\mathbf{r_1})]\,\mathrm{tr}[\Gamma^O(\mathbf{r_2},\mathbf{r_2})]} = \frac{\langle \Delta I(\mathbf{r_1})\Delta I(\mathbf{r_2})\rangle}{\langle I(\mathbf{r_1})\rangle\langle I(\mathbf{r_2})\rangle} \quad (2)$$

where '< >' denotes the ensemble average, $\gamma$ is the degree of coherence (DoC), $\Gamma^O(\mathbf{r_1}, \mathbf{r_2})$ is the 2 x 2 CP matrix of the random field, $\mathbf{E_O}(\mathbf{r}, t)$ and $\Delta I(\mathbf{r}) = I(\mathbf{r}) - \langle I(\mathbf{r}) \rangle$ is the spatial intensity fluctuation from its mean value. Numerator in Eq. (2) $\Delta I(\mathbf{r_1})\Delta I(\mathbf{r_2}) = C(\mathbf{r_1}, \mathbf{r_2})$ represents the cross-covariance of the random field. The degree of polarization (DoP) $P(\mathbf{r})$, which describes the spatial polarization distribution of the random field, can be calculated from the following relation

$$P^2(\mathbf{r}) = 2\gamma^2(\mathbf{r}, \mathbf{r}) - 1 \quad (3)$$

The maximum DoC and DoP for a spatially uniformly polarized random field is unity, whereas the maximum DoC for a spatially randomly polarized or depolarized field is 0.7 denoting the DoP is zero. The CP matrix for the field, $\mathbf{E_O}(\mathbf{r}, t)$ is written as

$$\Gamma^O(\mathbf{r_1}, \mathbf{r_2}) = \begin{bmatrix} \langle E_{Ox}{}^*(\mathbf{r_1})E_{Ox}(\mathbf{r_2})\rangle & \langle E_{Ox}{}^*(\mathbf{r_1})E_{Oy}(\mathbf{r_2})\rangle \\ \langle E_{Oy}{}^*(\mathbf{r_1})E_{Ox}(\mathbf{r_2})\rangle & \langle E_{Oy}{}^*(\mathbf{r_1})E_{Oy}(\mathbf{r_2})\rangle \end{bmatrix} \quad (4)$$

After passing through a polarizer with its transmission axis oriented at an angle, $\theta$ with the x-axis, the field, $\mathbf{E_O}(\mathbf{r}, t)$ is modified as

$$\mathbf{E_P}(\mathbf{r}) = \left[\cos^2\theta\, E_{Ox}(\mathbf{r}) + \sin\theta\cos\theta\, E_{Oy}(\mathbf{r})\right]\hat{\mathbf{x}} + \left[\sin\theta\cos\theta\, E_{Ox}(\mathbf{r}) + \sin^2\theta\, E_{Oy}(\mathbf{r})\right]\hat{\mathbf{y}} \quad (5)$$

The CP matrix for the field, $\mathbf{E_P}(\mathbf{r})$ can be written following Ref. [22] as

$$\Gamma^P(\mathbf{r_1}, \mathbf{r_2}) = \begin{bmatrix} \Gamma'_{xx}(\mathbf{r_1}, \mathbf{r_2}) & \Gamma'_{xy}(\mathbf{r_1}, \mathbf{r_2}) \\ \Gamma'_{yx}(\mathbf{r_1}, \mathbf{r_2}) & \Gamma'_{yy}(\mathbf{r_1}, \mathbf{r_2}) \end{bmatrix} \quad (6)$$

where
$\Gamma'_{xx} = d\,\Gamma^O_{xx} + a\,\Gamma^O_{xy} + a\,\Gamma^O_{yx} + c\,\Gamma^O_{yy}$;
$\Gamma'_{xy} = a\,\Gamma^O_{xx} + c\,\Gamma^O_{xy} + c\,\Gamma^O_{yx} + b\,\Gamma^O_{yy}$;
$\Gamma'_{yx} = a\,\Gamma^O_{xx} + c\,\Gamma^O_{xy} + c\,\Gamma^O_{yx} + b\,\Gamma^O_{yy}$;
$\Gamma'_{yy} = c\,\Gamma^O_{xx} + b\,\Gamma^O_{xy} + b\,\Gamma^O_{yx} + e\,\Gamma^O_{yy}$;
and $a = \sin\theta\cos^3\theta$, $b = \sin^3\theta\cos\theta$, $c = \sin^2\theta\cos^2\theta$, $d = \cos^4\theta$ and $e = \sin^4\theta$.

The passage of the spatially randomly polarized field through a polarizer makes the field spatially uniformly polarized. And the maximum value of the DoC and DoP calculated using the CP matrix, $\Gamma^P(\mathbf{r}, \mathbf{r})$ will always be fixed for any orientation of the polarizer. Hence it is not possible to make controlled modulation in the spatial polarization distribution using a single random field. In order to achieve the controlled modulation of the spatial polarization distribution, in the present work, the field $\mathbf{E_P}(\mathbf{r}, t)$ is superposed with another spatially uniformly polarized random field $\mathbf{E_R}(\mathbf{r}, t)$, referred as the reference random field with fixed polarization.

Considering that the electric field vector of the reference random field makes an angle $\phi$ with the x-axis, the field $\mathbf{E_R}(\mathbf{r}, t)$ can be written as

$$\mathbf{E_R}(\mathbf{r}, t) = E_R(\mathbf{r}, t) \cos\phi\,\hat{\mathbf{x}} + E_R(\mathbf{r}, t) \sin\phi\,\hat{\mathbf{y}} \quad (7)$$

where $E_R(\mathbf{r}, t)$ is the magnitude of the reference random field. The CP matrix for the field $\mathbf{E_R}(\mathbf{r}, t)$ can be written as

$$\Gamma^R(\mathbf{r_1}, \mathbf{r_2}) = \begin{bmatrix} \cos^2\phi\,\Gamma_R(\mathbf{r_1}, \mathbf{r_2}) & \sin\phi\cos\phi\,\Gamma_R(\mathbf{r_1}, \mathbf{r_2}) \\ \sin\phi\cos\phi\,\Gamma_R(\mathbf{r_1}, \mathbf{r_2}) & \sin^2\phi\,\Gamma_R(\mathbf{r_1}, \mathbf{r_2}) \end{bmatrix} \quad (8)$$

As the object and the reference random fields are experimentally generated from two independent scattering media, following Ref. [21], the resultant CP matrix can be written as sum of the CP matrix of the two individual fields as

$$\Gamma^T(\mathbf{r_1}, \mathbf{r_2}) = \Gamma^P(\mathbf{r_1}, \mathbf{r_2}) + \Gamma^R(\mathbf{r_1}, \mathbf{r_2}) \quad (9)$$

As the maximum DoC and DoP describe the spatial coherence of the random field at the same point and polarization distribution of the field, respectively, the calculation is focused at $\mathbf{r_1} = \mathbf{r_2}$ to study the changes in the polarization distribution of the resultant random field. Following the conditions for complete depolarization (DoP = 0) of the object random field i.e. $\Gamma^O_{ij}(\mathbf{r}, \mathbf{r}) = \langle E_{Oi}{}^*(\mathbf{r})\,E_{Oj}(\mathbf{r})\rangle = 0$ for $i \neq j$ and $|E_{Ox}(\mathbf{r})|^2 = |E_{Oy}(\mathbf{r})|^2$, and using $\Gamma^O_{xx}(\mathbf{r}, \mathbf{r}) = \Gamma^O_{yy}(\mathbf{r}, \mathbf{r}) = \Gamma_O(\mathbf{r}, \mathbf{r})$, it can be shown that the elements of the matrix, $\Gamma^P(\mathbf{r_1}, \mathbf{r_2})$ at $\mathbf{r_1} = \mathbf{r_2} = \mathbf{r}$ are modified as

$\Gamma'_{xx}(\mathbf{r}, \mathbf{r}) = (d + c)\,\Gamma_O(\mathbf{r}, \mathbf{r});\ \Gamma'_{xy}(\mathbf{r}, \mathbf{r}) = (a + b)\,\Gamma_O(\mathbf{r}, \mathbf{r});$
$\Gamma'_{yx}(\mathbf{r}, \mathbf{r}) = (a + b)\,\Gamma_O(\mathbf{r}, \mathbf{r});\ \Gamma'_{yy}(\mathbf{r}, \mathbf{r}) = (c + e)\,\Gamma_O(\mathbf{r}, \mathbf{r});$

It can be found that at $\mathbf{r_1} = \mathbf{r_2} = \mathbf{r}$, the $\Gamma_O(\mathbf{r},\mathbf{r}) = \langle I_O \rangle$ and similarly $\Gamma_R(\mathbf{r},\mathbf{r}) = \langle I_R \rangle$. Assuming spatial stationarity and ergodicity of the random field, the ensemble average can be replaced with space average. If the average spatial intensity of the object and reference random field can be made equal i.e. $\langle I_O \rangle = \langle I_R \rangle$, and considering $\Gamma_O(\mathbf{r},\mathbf{r}) = \Gamma_R(\mathbf{r},\mathbf{r}) = \Gamma_N$, the resultant CP matrix at same point i.e. polarization matrix, $\Gamma^T(\mathbf{r},\mathbf{r})$ in Eq. (9) is modified using triangular relation as

$$\Gamma^T(\mathbf{r},\mathbf{r}) = \begin{bmatrix} \cos^2\theta + \cos^2\phi & \sin\theta\cos\theta + \sin\phi\cos\phi \\ \sin\theta\cos\theta + \sin\phi\cos\phi & \sin^2\theta + \sin^2\phi \end{bmatrix} \Gamma_N \quad (10)$$

The square of the maximum DoC can be calculated as a function of $\theta$ using Eq. (10) and is given as

$$\gamma^2(\mathbf{r},\mathbf{r}) = \frac{tr[\Gamma^T(\mathbf{r},\mathbf{r})\Gamma^{T\dagger}(\mathbf{r},\mathbf{r})]}{|tr[\Gamma^T(\mathbf{r},\mathbf{r})]|^2} = \frac{(\cos^2\theta+\cos^2\phi)^2 + 2(\sin\theta\cos\theta+\sin\phi\cos\phi)^2 + (\sin^2\theta+\sin^2\phi)^2}{(\cos^2\theta+\cos^2\phi+\sin^2\theta+\sin^2\phi)^2} \quad (11)$$

The maximum DoC, $\gamma(\mathbf{r},\mathbf{r})$ can be simplified as

$$\gamma(\mathbf{r},\mathbf{r}) = \sqrt{\frac{1}{2}[1 + \cos^2(\theta \sim \phi)]} \quad (12)$$

The DoP, $P(\mathbf{r})$ can be calculated using Eq. (12) as

$$P(\mathbf{r}) = |\cos(\theta \sim \phi)| \quad (13)$$

It is observed from Eq. (12) and (13) that the maximum DoC and DoP can be modulated sinusoidally, which infers the control of single point correlation and spatial polarization distribution of the random field. It is also observed that the spatial polarization distribution of the resultant random field can be tuned from a uniformly polarized case to a randomly polarized case by controlling the polarizations of the constituent random fields. In a similar fashion, from the study of DoP, angle between the polarization vectors of two random fields can also be determined. The experimental demonstrations of the controlled modulation of the maximum DoC and DoP for a horizontally polarized reference random field ($\phi = 0$) are discussed below.

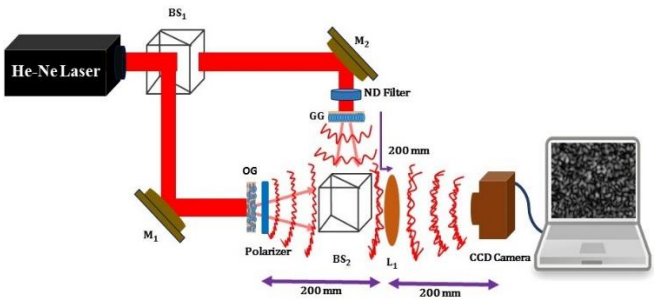

Fig. 1. The schematic diagram of the experimental set-up (color online)

The schematic of the experimental set-up for the present study is shown in Fig. 1. A horizontally polarized laser beam of 632.8 nm wavelength, from a He-Ne laser source, enters a Mach-Zehnder interferometer formed by beam splitters $BS_1$, $BS_2$ and mirrors $M_1$, $M_2$. The beam reflected from $BS_1$ gets folded by mirror $M_1$ and passes through a multiple scattering medium, here an opal glass (OG) plate. The speckle generated from OG is referred as object speckle. The beam transmitted through $BS_1$ gets folded by mirror $M_2$ and propagates down to another random scattering medium, a ground glass (GG) plate. The speckle generated from GG is referred as reference speckle. The intensity of the reference speckle is controlled using a neutral density (ND) filter to satisfy the condition mentioned before Eq. (10). The far-field superposition of the speckles generated from the OG and GG plates are recorded by a CCD camera placed at the back focal plane of a Fourier transforming lens, $L_1$ of focal length 200 mm as shown in Fig. 1. As explained after Eq. (6), the object speckle is filtered using a polarizer and the superposed speckles are recorded for different orientations of the polarizer from $0^0$ to $360^0$ in steps of $10^0$. The object and reference speckles are also recorded separately for their characterization.

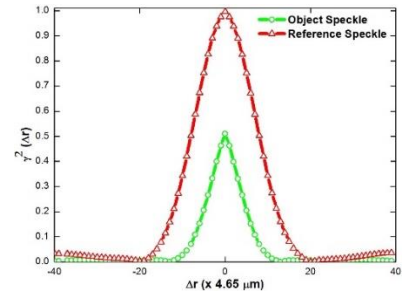

Fig. 2. The intensity correlation functions of the object and reference speckle; $\Delta\mathbf{r} = \mathbf{r_2} - \mathbf{r_1}$ (color online)

The spatial coherence and polarization property of the recorded speckles are studied using intensity correlation following Eq. (2) and (3), respectively. The intensity correlation functions of the object and reference speckle are presented in Fig. 2. It can be concluded from the maximum values of $\gamma^2(\Delta\mathbf{r})$ and using Eq. (3), that the reference speckle is fully spatially polarized, whereas the object speckle is completely depolarized. The depolarization in the object speckle is observed in the present experimental configuration because of the multiple scattering of light inside the OG plate. Depolarized speckle can also be generated by coherently mixing two orthogonal polarization components having completely random independent phases. This can be realized by considering two orthogonal polarization components coming from two different portions of the ground glass [17]. Random scattering from a birefringent scatterer can also be utilized to generate spatially depolarized speckle [22].

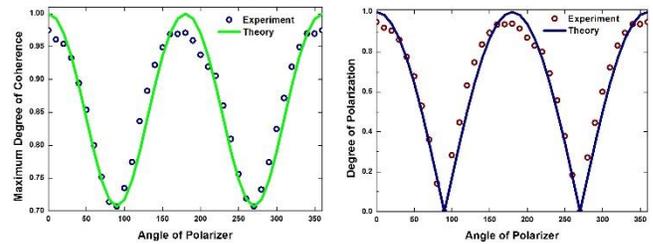

Fig. 3. The experimental results (open circles) and the theoretical prediction (solid line) on controlled modulation of the maximum DoC and DoP, for object speckle generated using OG plate (color online)

The experimental results (open circles) along with the theoretical prediction (solid line) on controlled modulation of the coherence and depolarization of the resultant speckle are presented in Fig. 3. It can be observed from Fig. 3 that the maximum DoC and DoP is changing sinusoidally as a function of the orientation of the transmission axis of the polarizer. It can also be observed that the spatial polarization distribution of the resultant speckle can be tuned from a uniformly polarized case (DoP = 1) to a completely depolarized case (DoP = 0) by proper choice of polarizations of the constituent speckles. The slight deviation of the experimental result from the theoretical prediction observed in Fig. 3 is due to leakage in the polarizer which is confirmed experimentally (not shown here) and also because of the difference in average intensity of the x and y component of the depolarized speckle due to multiple scattering. To confirm the effect of the average intensity difference on the deviation of the experimental results from the theory, the experiment is also carried out by replacing the OG plate in Fig. 1 with a balanced polarization interferometer of Mach-Zehnder type with a GG plate at its output (the experimental set up is not shown here). In this case, the depolarized speckle is generated due to the superposition of two spatially displaced, orthogonally polarized beams of equal intensity at the output of the polarization interferometer that are illuminated on the GG plate. The experimental results on the modulation of the maximum DoC and DoP obtained in this case are shown in Fig. 4, which confirm our assumption.

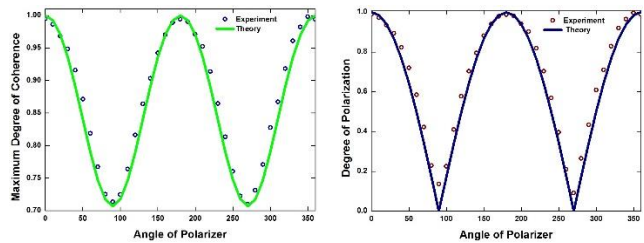

Fig. 4. The experimental results (open circles) and the theoretical prediction (solid line) on controlled modulation of the maximum DoC and DoP, for object speckle generated using GG plate (color online)

In conclusion, we have proposed and demonstrated a simple technique without using any specific diffracting element to achieve sinusoidal modulation of the maximum degree of coherence and the degree of polarization. It is also shown that the spatial polarization distribution of the resultant speckle can be modulated from a uniformly polarized case to a completely depolarized case by controlling the mutual angle between the polarization vectors of the constituent speckles. The proposed technique can find application in sensing and biomedical studies, where the rotation of the electric field vector can be determined by mixing it with another uniformly polarized scattered field. Moreover, similar study can also be extended to the temporal domain.

**Funding.** MMB acknowledges support from SERB-DST (Project No. SR/FTP/PS-170/2013). RKS acknowledges the support from SERB-DST (Project No. EMR/2015/001613).